\documentclass[pra,twocolumn,showpacs]{revtex4}
\usepackage{graphicx}
\usepackage{hyperref}
\usepackage{amsmath}

\def\id{{\mathchoice {\rm 1\mskip-4mu l} {\rm 1\mskip-4mu l} {\rm
1\mskip-4.5mu l} {\rm 1\mskip-5mu l}}}
\def\ket#1{| #1 \rangle}

\def\ketbra#1#2{| #1 \rangle\!\langle #2 |}

\newcommand {\be} {\begin{eqnarray}}
\newcommand {\ee} {\end{eqnarray}}
\newcommand{\ham}{\mathcal{\hat{H}}}

\begin{document}

\title{Experimental Implementation of Discrete Time Quantum Random Walk \\ on an  NMR Quantum Information Processor }

\author{C.A. Ryan\footnote{email address : c4ryan@iqc.ca}$^\dagger$}
\author{M. Laforest\footnote{These authors contributed equally to this work.}}
\author{J.C. Boileau}
\author{R. Laflamme}
\affiliation{Institute for Quantum Computing, University of Waterloo, Waterloo, ON, N2L 3G1, Canada.}

\date{\today}
\begin{abstract}
We present  an experimental implementation of the coined discrete time quantum walk on a square using a three qubit liquid state nuclear magnetic resonance (NMR) quantum information processor (QIP).  Contrary to its classical counterpart, we observe complete interference after certain steps and a periodicity in the evolution.  Complete state tomography has been performed for each of the eight steps making a full period.  The results have extremely high fidelity with the expected states and show clearly the effects of quantum interference in the walk.   We also show and discuss the importance of choosing a molecule with a natural Hamiltonian well suited to NMR QIP by implementing the same algorithm on a second molecule.  Finally, we show experimentally that decoherence after each step makes the statistics of the quantum walk tend to that of the classical random walk.  
\end{abstract}


\maketitle

\section{Introduction}
The idea of exploiting the quantum mechanical behaviour of a device to gain power in simulating quantum systems was first introduced by Richard Feynman \cite{Fey82a}. The field of quantum computing has since grown enormously with the discovery  of  two algorithmic pillars: Shor's factoring algorithm \cite{Sho94a} and Grover's search algorithm \cite{Gro97a}.  Both of these demonstrate a clear speed-up over their classical counterparts.  Following in this path, many other quantum algorithms have been developed that provide a speed-up \cite{Kit95a,Sim97a,EJ98a}.  A more recent addition to the family of quantum algorithms which demonstrate an exponential speed-up,  are those based on the quantum random walk  -  the quantum version of the successful classical random walk \cite{CCD+02a}.

There is however, a need to explore more than the simple computational properties of the algorithms.  They must also be experimentally tested in real devices and their relative ease of implementation compared and considered.  In particular, in QIP devices where we are controlling the natural Hamiltonian, it is important to choose a system where the Hamiltonian is amenable to automatic and systematic control.  This can be explored by implementing the same algorithm in different molecules and contrasting the performance.     Although many different implementation schemes have been proposed for the quantum random walk algorithm, using for example trapped ions \cite{TM02a}, an optical lattice \cite{DRKB02a}, cavity QED \cite{SBTK03a}, or an optical cavity \cite{KRS03a}, these have not been tested.  The only experimental test of a quantum walk is the continuous time version of a quantum walk on a square using a two qubit  nuclear magnetic resonance (NMR) quantum information processor (QIP)  \cite{DLX+03a}.  This work showed the contrast between a classical and quantum random walk and showed the influence of entanglement on the probability distribution of the quantum walk.  Here, we present an experimental proof of principle experiment of a discrete time quantum walk on a square.  The effects of decoherence on the quantum random walk has been investigated by several authors and indeed, it may offer some benefits \cite{KT02a,BCA02b}.  Therefore, we also explored the quantum to classical transition of our walk under the addition of decoherence to the coin.  Furthermore, we compared and contrasted two different control schemes and molecules by implementing the algorithm on two molecules. 

\section{Quantum Random Walks}
In the development of deterministic classical randomized algorithms, the methods of Markov chains and random walks have played a fundamental role \cite{MR95a}.  These algorithms can be divided into two categories:  continuous time random walks when the walker has a probability per unit time to make a move; and, discrete time random walks where the walker moves at defined time-steps.   Since these processes are stochastic, it is not surprising that they have quantum counterparts.  The quantum versions however, show  remarkable differences with their classical analogues.  The continuous time quantum walk (CTQW)\cite{FG98b} has been shown to provide an exponential speed-up in propagation through a graph \cite{CFG02a,CCD+02a}.  The discrete time quantum walk (DTQW) \cite{ADZ93a}  plays an important role in the speed up of a quantum algorithm design for spatial searching \cite{SKW03a,AA03a,AKR05a}.

One step of a classical discrete time random walk on a circle with $n$ nodes, denoted by $\{0, \ldots,  n-1\}$, is performed by repetition of the following two steps:  (1) the walker first flips a coin and then  (2) moves either clockwise or counterclockwise depending on the outcome of the coin toss. 

\begin{figure*}[ht]
\includegraphics[angle=90,scale = 0.525]{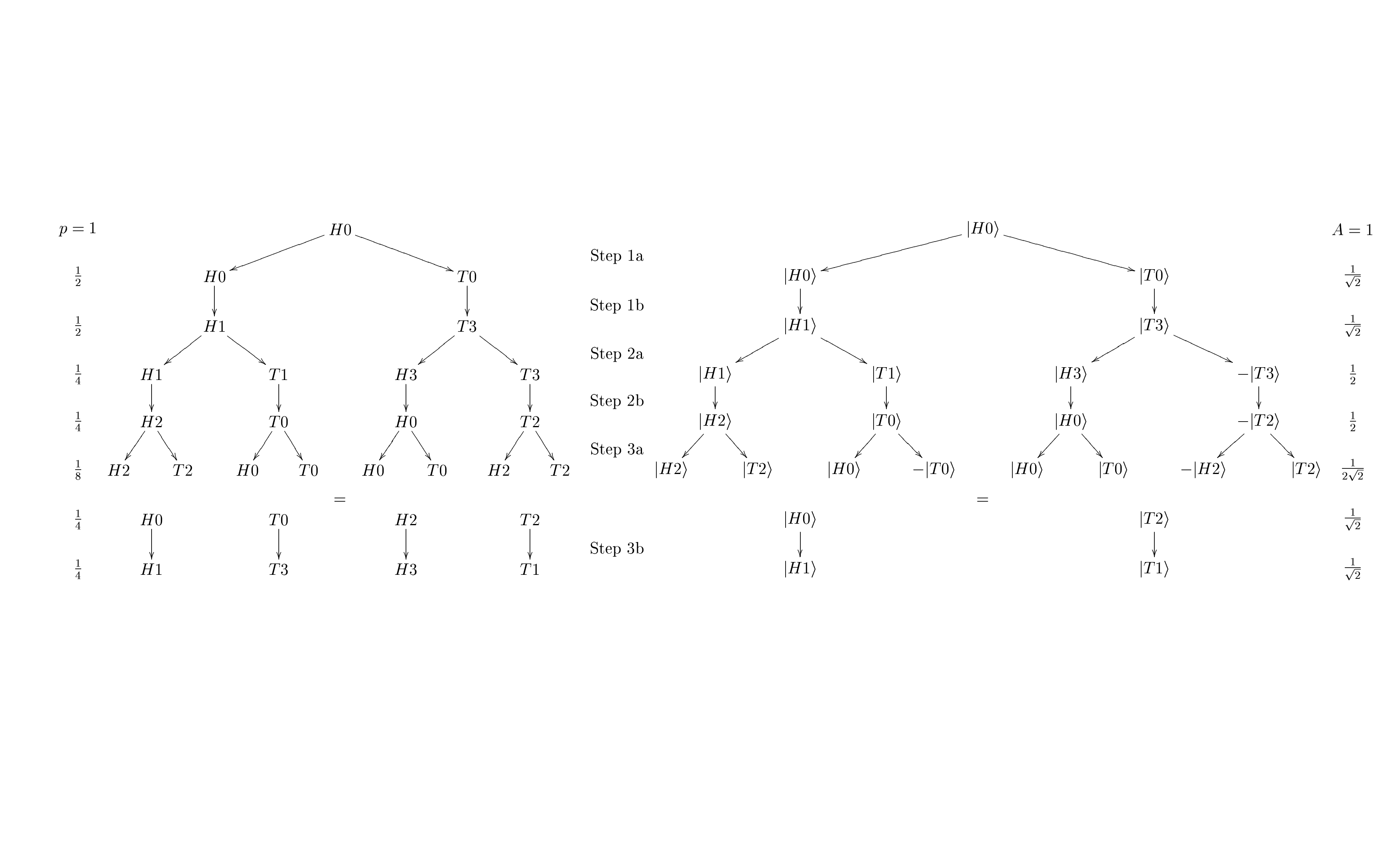}
\caption{\label{bothdynamics}
Comparison of the dynamics of the classical (left) and quantum (right) random walk on a square for three steps.  $H$ or $T$  represents the state of the coin and the number, the position of the walker at one of the four nodes of the square. $p$ is the probability of each classical state and A is the probability amplitude for the quantum state.  Part (a) for each step is the coin flip and part (b) the movement around the square.  In both cases the walker starts at node 0 with the coin in the heads state.   After one step he has a fifty percent probability of being at either of nodes 1 or 3.  Then, in the second step he goes to either 0 or 2 with fifty percent probability.   In the third step however, the two types of walk diverge.  The classical walk continues to oscillate and the probability remains spread out.  In the quantum walk on the other hand, the probabilities interfere and cancel out leaving all the probability in one corner after three steps. }
\end{figure*}

If we perform a quantum mechanical treatment of the situation, we can label the nodes with a mutually orthonormal set of state vectors $\{\ket{i}\}_{i=0}^{n-1}$.  The coined DTQW on the circle  can be seen as ``quantumly'' flipping a coin degree of freedom using a unitary operation and then coherently  moving the walker position degree of freedom clockwise, or counterclockwise, conditioned on the state of the coin \cite{AAKV01a}.   For a Hadamard walk, the coin flipping operation is simply the Hadamard gate described by the matrix, 
\be
\hat{H}&=&\frac{1}{\sqrt{2}}\begin{pmatrix}1&1\\1&-1\end{pmatrix}.
\ee

Now, the conditional shift operator is defined as
\be
\hat{S}\ket{H}\ket{i} &=&\ket{H}\ket{i\ominus 1} \label{shift1}\\
\hat{S}\ket{T}\ket{i} &=&\ket{T}\ket{i\oplus 1\label{shift2}} ,
\ee
where $\oplus$ and $\ominus$ are understood to be addition and subtraction modulo $n$ and $\ket{H}$ and $\ket{T}$ describe the two basis states of the coin. Therefore, if the walker is in position $\ket{i}$, he will move clockwise to the position $\ket{i\ominus 1}$ if the coin in the state $\ket{H}$, or counterclockwise to $\ket{i\oplus 1}$ if the coin in the state $\ket{T}$. We can write this operator as
\be
\hat{S}&=&\sum_{i=0}^{n-1}\left(\ketbra{H}{H}\otimes\ketbra{i\ominus1}{i}+\ketbra{T}{T}\otimes\ketbra{i\oplus1}{i}\right).
\label{shifttotal}
\ee

Then, one step of the DTQW is defined as applying the operator
\be
\label{unistep}
\hat{W}&=&\hat{S}(\hat{H}\otimes\id).
\ee

On a circle, this type of algorithm shows destructive interference effects and a probability distribution that is periodic in time.  The contrasting dynamics for the classical and quantum random walks are shown in Fig. \ref{bothdynamics}.   As opposed to the classical walk where the probability is always spread out, the quantum walk has steps where the probability amplitudes interfere such that all the probability comes back to one node.   Furthermore, this walk is periodic in that  after eight steps, the corresponding propagator is equal to the identity and the system comes back to its original state.  

\begin{figure}[hbp]
\includegraphics[scale=0.22]{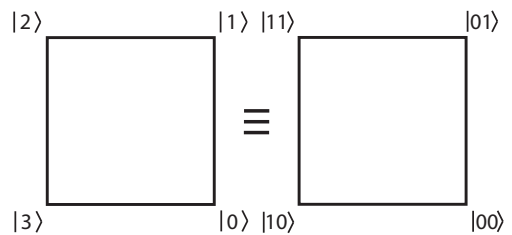}
\caption{\label{square} Logical labeling of the nodes on which we implemented the DTQW.  With this labeling, flipping the first qubit corresponds to a horizontal move and flipping the second qubit, a vertical move. }
\end{figure}

In our experimental setup we have three qubits available, which allows one qubit to describe the coin state and two for the position state.  Thus, we have $n=4$, and we are performing a discrete quantum walk on a square.  The shift operator defined in Eq. \ref{shifttotal} would require a complicated quantum circuit involving a Toffoli gate.   We can simplify the circuit required by using a  shifting operator that moves the walker along a direction vector, i.e. horizontally or vertically (this also is analogous to the random walk on the hypercube \cite{MR02a}).  Therefore, if we label the corners of the square as shown in Fig \ref{square}, the shift operator on the three qubit register becomes,

\be
\label{unimove}
\hat{S}&=&\hat{P}_H\hat{X}^2+\hat{P}_T\hat{X}^3 \nonumber\\
&=&(\hat{P}_H\hat{X}^2+\hat{P}_T)(\hat{P}_T \hat{X}^3+\hat{P}_H)\nonumber\\
&=&(\hat{X}^1Cnot^{1,2}\hat{X}^1)Cnot^{1,3},
\ee
with  $X$ denoting the standard $\sigma_x$ Pauli matrix, $P_{H/T}$ are the projectors on the two coin states, and the superscript indicates on which of the qubits the action is performed.  Here, it is understood that the first qubit represents the coin and the second and third, the position register.  The resulting probabilities for each step are shown in Table \ref{probtable}. 

\begin{table}[htbp]
\begin{tabular}{||c||c|c|c|c||c|c|c|c||}
\hline\hline
&\multicolumn{4}{c||}{Classical}&\multicolumn{4}{c||}{Quantum} \\
\hline
Corner &0&1&2&3&0&1&2&3 \\
\hline\hline
Step 0&1		     &---		&	---	     &	 	---	&	1	     &		---	& ---		    &	---\\
\hline 
Step 1& ---     & 0.5 &	---	     &	 	0.5	&	---	     &		0.5	& --- &0.5\\
\hline
Step 2 & 0.5	    & ---		&	0.5	     &		---	&	0.5	     &		---	&	0.5	    & --- \\
\hline
Step 3 & ---	   &  0.5		& --- & 0.5     &	---     	     &		---	& --- & 1 \\
\hline		    
Step 4& 0.5 	  & ---		&	0.5	     &		---	&	---	     &		---	&	1	    & --- \\
\hline		 
Step 5&---		     & 0.5		& --- & 0.5	& ---	     &0.5 & ---		   &	0.5 \\
\hline
Step 6 & 0.5    & ---		& 0.5	     & ---	&	0.5	     &		---	&	0.5	    &  --- \\
\hline
Step 7  & ---      & 0.5 &	---	    &	 0.5	&	---	      & 1 & ---		   &	---\\
\hline
Step 8& 0.5     &---		& 0.5	     &	 	---	& 1	     &  --- 	&---		    &	---\\
\hline
\hline
\end{tabular}
 \caption{\label{probtable}
 Probability to be in each of the corner states as denoted in Fig. \ref{square}.  While in the classical random walk the probability always remains spread out between two corners, in the quantum random walk all the probability returns to one corner at certain time steps.}
\end{table}

\section{Liquid state NMR quantum information processing}
\subsection{The basic principles}
A liquid state NMR QIP consists of an ensemble of roughly $10^{20}$ identical molecules dissolved in a liquid solvent.  Due to the fast tumbling motion of the molecules, they are essentially decoupled from each other; ideally all the molecules have the same evolution.  We can think of the quantum register made of qubits that correspond to the spin $\frac{1}{2}$ nuclei within each molecule.   The sample is placed in a strong homogeneous magnetic field which provides the quantization axis and causes the spins to precess around the axis of the field.  It is possible to implement single qubit gates using radio-frequency (r.f.) pulses resonant with the precession frequency, which can effect a rotation about any axis orthogonal to the axis of the external field.  Two qubit gates are effected through the coupling from the natural Hamiltonian, which produces a controlled phase gate\cite{LKC+02a}.

 If the molecule used contains $n$ distinguishable nuclei and the magnetic field is aligned along the z-axis, then the system Hamiltonian is approximated by,
 \be
 \ham=\pi\sum_{i=1}^{n}\nu_i\hat{Z}_i+\frac{\pi}{2}\sum_{i>j}J_{ij}\hat{Z}_i\otimes \hat{Z}_j,
 \ee
where $\nu_i$ is the Larmor frequency of spin $i$ in Hz, $J_{ij}$ is the coupling strength between spin $i$ and $j$ in Hz and $Z$ in the conventional Pauli operator $\sigma_z$.  The interaction part of the Hamiltonian can be approximated to the above Ising form (weak coupling regime or secular approximation) only if the difference between any two nuclei Larmor frequencies is much greater than the coupling between the nuclei.  We can also turn off the coupling between any two spins as needed by applying refocussing r.f. pulses.
 
\subsection{Implementing dephasing in NMR}\label{gradient}
We can apply a controllable amount of decoherence to selected spins using gradient techniques in NMR. Consider only one nucleus with state $\rho$ and suppose we work in the rotating frame of that spin.  On a NMR spectrometer, it is possible to apply a gradient to the external magnetic field.  During the time that the gradient is applied, the spins will precess at different frequencies depending on their position in the sample.  The state of the ensemble will then be given by an average over the observable sample,
\be
\rho'&=&\frac{1}{2a}\int_{-a}^{a}e^{-i(\alpha'\gamma tz/2)\hat{Z}}\rho e^{i(\alpha'\gamma tz/2)\hat{Z}}dz,
\ee
where $2a$ is the length of the sample, $t$ is the interval of time the gradient is being applied and $\alpha'=\alpha/\hbar$ and $\gamma=\nu/B_z$, the gyro-magnetic ratio of the nucleus.  If we compute the integral, it can be shown that,
\be\label{dephasing}
\rho'&=&(1-p)\rho+pZ\rho Z, \nonumber\\
&&p=\frac{1}{2}\left(1-\frac{1}{\alpha'\gamma ta}\sin{(\alpha'\gamma ta)}\right),
\ee
which is the exact form of a $z$-dephasing decoherence. The amount of dephasing can be controlled by the strength and time of the gradient pulse.  Particular spins can be protected from the applied decoherence by applying a 180 degree rotation and applying a second gradient of the same strength and time.  This second gradient will reverse  the dephasing of the rotated spins and double it on the spins that were not rotated.

\section{The experiment}
We implemented the quantum walk algorithm on two molecules: trans-crotonic acid and trichloroethylne (TCE).  This allowed us to compare the quality of two different methods of control and the merits of the two molecules.  

\subsection{Implementation on crotonic acid}
\begin{figure}[htb]
\includegraphics[scale=1]{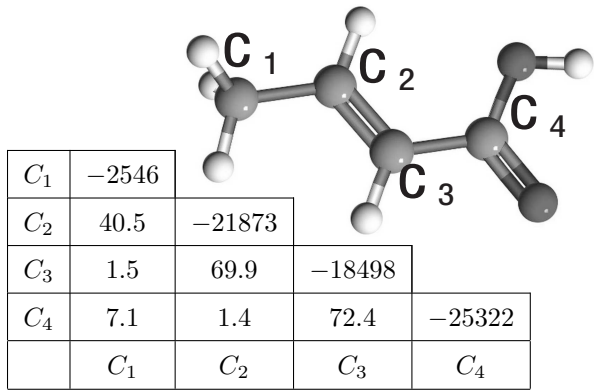}
\caption{Molecular structure of trans-crotonic acid and its Hamiltonian parameters.  The chemical shifts are given as the diagonal elements and the coupling strength (Hz) by the off-diagonal elements.  Note that since the darkly shaded unlabeled nuclei are oxygen whose natural abundance of $^{16}O$ with 0 spin is close to 100 $\%$.  Therefore, the two oxygen nuclei do not couple with the rest of the molecule and can be ignored.  Lightly shaded unlabeled nuclei are hydrogen which were decoupled during the experiment.  \label{crot}}
\end{figure}

The seven qubit molecule trans-crotonic acid (four carbons, two hydrogens and one methyl group) has  been used in experimental demonstrations of quantum algorithms, such as  quantum error correction \cite{KLMN01a,BVFC04a}, and quantum simulations\cite{NSO+04a}.  In this experiment, we used the carbon back-bone of labeled trans-crotonic acid in a solution of deuterated acetone.    The hydrogen nuclei were decoupled using standard heteronuclear decoupling techniques\cite{SKF83a}.  We  used $C_3$ as the coin and $C_2$ and $C_4$ as the position register (see Fig. \ref{crot}).   $C_1$ was used as a labeling spin to ease the creation of the initial state.    On a Bruker DRX Avance 600 NMR spectrometer, the molecule has the Hamiltonian parameters shown in Fig. \ref{crot}.

Since our system is homonuclear, the control of individual qubits is achieved through soft gaussian-like r.f. pulses at the Larmor frequency of the target nucleus.   The length of the soft pulses is of the order of the inverse of the smallest chemical shift difference with the other nuclei.  In our experiment the length of the selective pulses on $C_1$ and $C_2$,$C_3$,$C_4$ were $192\mu$s and $704\mu$s respectively.

\subsubsection{Initial state preparation}

The experiment required  the initial state
\be
\rho_{in}=*\otimes\ketbra{000}{000}=*(\id+\hat{Z})(\id+\hat{Z})(\id+\hat{Z}).
\ee
We created the labeled pseudo-pure state $Z000$ (using the notation $C_1C_2C_3C_4$) following the spatial averaging technique elaborated in \cite{KLMT00a}.  
 
\subsubsection{Pulse sequence implementation} 
\begin{figure}[htb]
\includegraphics[scale=0.35]{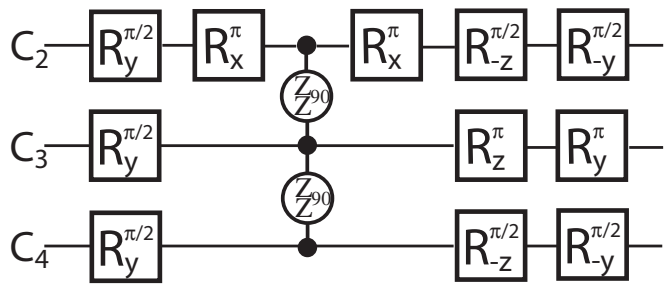}
\caption{NMR pulse sequence representing one step of DTQW.  The notation $R_i^{\theta}$ means a rotation of an angle $\theta$ around the axis $i$.  Refocussing pulses are not shown.  Since each nucleus is tracked in its own rotating frame, rotations about the z-axis are implemented instantaneously through a change of reference frame.   \label{pulseseq}}
\end{figure}

The unitary of one step of the DTQW from Eq \ref{unimove} was translated to a sequence of pulses and coupling gates as shown in Fig. \ref{pulseseq}.   Although many pulse sequences are possible through the use of commutation rules, this particular one was designed to be the most efficient due to the cancellations possible during multiple step sequences.  Moreover, the $ZZ$ gates are achieved simultaneously, which shortens the overall pulse sequence, thus reducing decoherence effects.  Commutation rules were also used to cancel pulses between the final step and the readout pulses.    The ideal pulse sequence of rotations and couplings was then input into a pulse sequence compiler which numerically optimized the timing and phases of the pulses \cite{BJKL:05a}.  The compiler pre-simulates the selective r.f. pulses using an efficient pair-wise simulation and then decomposes the simulated unitary into phase and coupling errors sandwiching the ideal selective pulse.  These errors can then be taken into account by the refocussing scheme and phase of the pulses, so that the overall unitary is as close to the desired one as possible. 

Since we are concerned with the final state of only three qubits in this experiment, complete state tomography is still feasible.   On a three qubit system in NMR,  only seven different readout pulses are required to rotate each term of the density matrix into observable simple single coherences \footnote{Readout pulses yII,IIy,IIx,yyI,Ixx,yyy,xxx are sufficient}.  And, since we were operating on a homonuclear system, observing the signal from all spins in one experiment was possible, with some post-processing to adjust for the correct phase of each individual rotating frame.   The coupling between the labeling spin $C_1$ and the other three qubits is resolvable and so the presence of the labeling spin does not interfere with the tomography of $C_2, C_3$ and $C_4$.

\subsubsection{Experimental results}
For the state tomography each of the peaks in the spectra were fitted using absorption and dispersion Lorenzian peaks.  The full density matrix was then reconstructed by appropriately summing up the corresponding Pauli terms.  Where two experiments gave values for the same density matrix terms, the values were simply averaged.  As we observed only $C_2, C_3$ and $C_4$, the term $ZIII$ could not be determined.  A suitable amount of that term  was subsequently added to the density matrix so as to make the initial state as close to $Z000$ as possible.  This amount was then kept constant for the density matrix reconstruction in  subsequent experiments.

To quantify the success of our experiments, we computed the fidelity of the experimental density matrix to both the ideal and simulated results.   In NMR, all states are nearly completely mixed and the fidelity measure introduced in \cite{FPB+02a} is appropriate.  We can compare one density matrix to another using the following formula, 
\be
F^{A,B}&=&\frac{Tr(\rho^{A}\rho^{B})}{\sqrt{Tr((\rho^{A})^2)}\sqrt{(Tr((\rho^{B})^2)}}.
\ee

We made two comparisons.  Firstly we compare the experimentally determined density matrix to the theoretically expected result.  The theoretical result is achieved by multiplying the ideal initial state by the ideal propagator.  To investigate how well we understand our control of the system, we also compare the fidelity of the results from a simulation of the experiment to the theoretical result.   

\begin{table}[htbp]
 \begin{tabular}{||c|c|c||}
 \hline\hline
 & Experimental & Simulated \\
 \hline\hline
 Step0 &  $98\pm5$ & --- \\
 \hline
 Step1  & $97\pm5$ & 98 \\
 \hline
 Step2 & $98\pm5$ & 98  \\
 \hline
 Step3  &$92\pm5$ & 98 \\
 \hline
 Step4  & $99\pm5$ & 98 \\
 \hline
 Step5  & $94\pm5$ & 97 \\
 \hline
 Step6 & $96\pm5$ & 97  \\
 \hline
 Step7  & $96\pm5$ & 97 \\
 \hline
 Step8 & $87\pm4$ & 97  \\
 \hline\hline
 \end{tabular}
 \caption{Fidelities (in percent) of experimental and simulated results.  The first column gives the fidelity of experimental density matrix determined from the tomography, with respect to the theoretical expected density matrix.  The second column gives the fidelity of the simulation results.  Errors are estimated from the fitting procedure.  Note that since computer simulation of the spatial averaging that occurs during the pseudo-pure preparation is difficult and inaccurate, the initial state for the simulation was the experimental pseudo-pure state determined from the tomography.  The fluctuations observed in the fidelity come from uncertainties in the fit and instabilities in the spectrometer over the course of the experiment.   \label{results}}
 \end{table}

The fidelities of simulated and experimental results are compared in Table \ref{results}.  The loss of fidelity in our experiment, over and above that of the simulated control errors is explained from three sources which are not taken into account by the simulation.  We have losses from T2 relaxation.  Although our pulse sequence is short compared with the T2 relaxation times, during the quantum walk algorithm, the state is often in high coherences, which decay much faster than the simple T2 time. Inhomogeneities in the strong magnetic field also cause extra relaxation and dephasing.  Further losses come from inhomogeneities of the r.f. field used to implement rotations and pulse angle mis-calibration. 

\subsubsection{Addition of decoherence on the coin}

In a subsequent experiment, we added dephasing decoherence to the entire qubit register using the technique described in section \ref{gradient}.  We expect the behaviour of the quantum walk should converge to the classical walk as the decoherence becomes complete after each step.   To demonstrate this claim experimentally, we implemented the quantum random walk for four steps, adding decoherence of a certain strength between each step of the walk.  The differences between quantum and classical walk is manifested in the different probabilities of being in each of the corners  after each step.  The results are shown in table \ref{decoresults} for gradient strengths corresponding to no, partial and full decoherence. 

\begin{table*}[ht!]
\begin{center}
\begin{tabular}{||c||c|c|c|c||c|c|c|c||c|c|c|c||}
\hline
\hline
&\multicolumn{12}{c||}{Quantum Walk with Decoherence}\\ 
\cline{2-13}
&\multicolumn{4}{c||}{None} &\multicolumn{4}{c||}{Partial} &\multicolumn{4}{c||}{Full}\\ 
\hline
Corner &0&1&2&3&0&1&2&3&0&1&2&3 \\\hline
\hline
Step 0 & $100\pm8$ & $0\pm1$ & $4\pm1$ & $-2\pm1$ & $100\pm8$ & $0\pm1$ & $4\pm1$ & $-2\pm1$ & $100\pm8$ & $0\pm1$ & $4\pm1$ & $-2\pm1$ \\ 
 \hline 
Step 1 & $2\pm1$ & $57\pm4$ & $-1\pm1$ & $43\pm3$ & $2\pm1$ & $58\pm4$ & $-2\pm1$ & $44\pm3$ & $0\pm1$ & $59\pm4$ & $-2\pm1$ & $45\pm4$ \\ 
 \hline 
Step 2 & $57\pm4$ & $1\pm1$ & $44\pm3$ & $-1\pm1$ & $50\pm4$ & $7\pm1$ & $40\pm3$ & $5\pm1$ & $51\pm4$ & $4\pm1$ & $46\pm4$0 & $1\pm1$ \\ 
 \hline 
Step 3 & $7\pm1$ & $14\pm1$ & $3\pm1$ & $78\pm6$ & $2\pm1$ & $14\pm1$ & $1\pm1$ & $84\pm6$ & $-1\pm1$ & $53\pm4$ & $-3\pm1$ & $53\pm4$ \\ 
 \hline 
Step 4 & $15\pm1$ & $-1\pm1$ & $84\pm6$ & $3\pm1$ & $19\pm2$ & $1\pm1$  & $78\pm6$ & $4\pm1$ & $50\pm4$ & $0\pm1$ & $53\pm4$ & $-1\pm1$ \\ 
 \hline 
\hline
\end{tabular}
\end{center}
\caption{Classical versus estimate of quantum probability to be in each corner of the square for one through four steps (c.f. Table \ref{probtable}).  The quantum results were obtained for gradient strengths corresponding to no, partial and full decoherence.   The experimental results were obtained by reconstructing the density matrix using the same fitting software used before and then applying the position measurement projectors to the reconstructed density matrix.  \label{decoresults}}
 \end{table*}

The divergence between the classical and quantum walk shows most clearly in steps three and four.  Whether the walk is classical or quantum, steps one and two yield the same measurement probabilities for the position (however the quantum version with decoherence will have coherent superposition states).  Analyzing the data from step 3 and 4, one can see that the quantum interference present so clearly in the quantum walk with no decoherence, is less obvious as the amount of decoherence increases.  Instead of the probability all collecting in one corner it remains spread out between two opposite corners - the same as in the classical walk.

The probabilities even with zero gradient strength do not correspond perfectly to the ideal quantum walk.  We believe these errors come from two sources.   Because the gradient does not commute with any pulses we were not able to use commutation rules to reduce the number of pulses during multiple step experiments.   Furthermore, gradient methods are hampered by diffusion and multiple gradients may lead to a return of signal that was ``erased" by a previous gradient.

\subsection{Comparison with the TCE molecule}
\begin{figure}[htb]
\includegraphics[scale=1]{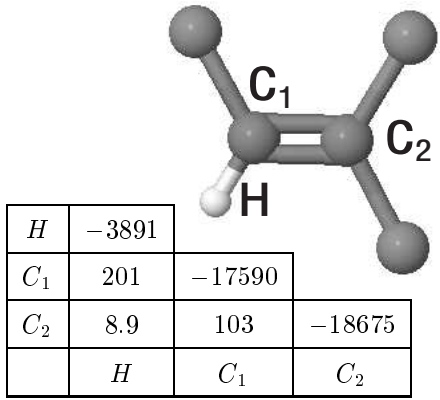}
\caption{Diagram of $^{13}C$ labeled TCE. The chemical shifts and couplings are given in the table.   Note that since the chlorine nuclei (unlabeled) have a spin of $\frac{3}{2}$, they have an electric quadrupole moment which causes them to decohere quickly and they have a very small coupling to the rest of the molecule which we can ignore in the natural Hamiltonian of the molecule. \label{TCE}}
\end{figure}

For comparison purposes and to show the importance of choosing a molecule with good characteristics in liquid state NMR quantum information processing, we show our results from our initial attempt to implement the quantum walk on the molecule tri-chloroethylene (TCE) - a molecule with which we have much less control due to the presence of strong coupling.  The molecule has been used for some initial demonstrations of quantum algorithms \cite{CMP+98a}.  A diagram of the molecule and the parameters of its Hamiltonian are shown in Fig. \ref{TCE}. 

\subsubsection{Pseudo-pure state preparation}
Since the TCE molecule contains only three qubits, we are unable to create the labeled pseudo-pure state that we used in the crotonic acid experiments.  Instead, we chose to use temporal averaging and add three separate experiments to achieve the initial state $\ket{000}$.  The three different initial states we used are,

\be
\rho_1&=&\hat{Z}\otimes(\id+\hat{Z})\otimes(\id+\hat{Z}) \nonumber \\
\rho_2&=&\id\otimes \hat{Z}\otimes(\id+\hat{Z}) \nonumber \\
\rho_3&=&\id\otimes\id\otimes(\id+\hat{Z}) 
\ee

If we add the results of these three experiments, it is equivalent to having performed the algorithm on the initial state
\be
\rho_{in}&=&\rho_1+\rho_2+\rho_3\nonumber \\
&=&(\id+Z)\otimes(\id+Z)\otimes(\id+Z)\nonumber \\
&=&\ketbra{000}{000}
\ee

Since there is only one hydrogen nucleus in the molecule, we can use broadband hard pulses to control it. One useful property of the TCE molecule in a 600Mhz spectrometer, is that the J-coupling between the two carbons is almost exactly 10.5 times smaller than the difference in chemical shift ($\Delta\nu$).  Therefore, during the time for a  $\frac{\pi}{2}\hat{Z}\hat{Z}$ coupling gate between the two carbons ($\Delta t=\frac{1}{2J_{C_1C_2}}$),  the relative chemical shift evolution of $C_2$ with respect to $C_1$ will be $\theta=2\pi\Delta\nu\Delta t=\frac{\pi\Delta\nu}{J_{C_1C_2}}=  -10.5\pi = -\pi/2$ mod$2\pi$.  Therefore, in the reference frame rotating at the Larmor frequency of $C_1$, every time there is a $\pi/2$ coupling between the carbons, an extra $R_{z}^{-\pi/2}$ is naturally performed on $C_2$.  

The chemical shift difference between the two carbons is small and the coupling  between them large, so selective pulses were impossible to achieve using the same technique of gaussian-shaped pulses used in the crotonic acid experiments.  These pulses  would be very long (roughly 5 ms) and the large coupling errors that would occur  during the pulse would be difficult to refocus.  Instead, it was possible to perform  selective pulses using hard pulses and the chemical shift evolution.   To illustrate the technique, we demonstrate how to perform a selective $\frac{\pi}{2}$ rotation of $C_2$.  If we use a reference frame rotating at the Larmor frequency of $C_1$, then, during a time $\tau=\frac{1}{4\Delta\nu}$, the spin $C_1$ will not precess  while $C_2$ will undergo a rotation of $-\pi/2$ around the $z$-axis.  Since, $\frac{1}{4\Delta\nu}$ is much less than the coupling time $\frac{1}{2J_{C_1C_2}}$, we can ignore the coupling between the two carbons and refocus only the hydrogen.   Using this selective $z$-rotation combined with hard pulses which rotate the two carbons together, we can perform a $\frac{\pi}{2}$ rotation with phase $\phi$ on only $C_2$ as follows:  
\be
(R_{\phi-\pi/2}^{\pi/2}&\otimes& R_{\phi-\pi/2}^{\pi/2})(\id\otimes R_z^{-\pi/2})(R_{\phi+\pi/2}^{\pi/2}\otimes R_{\phi+\pi/2}^{\pi/2})\nonumber\\
&=&\id\otimes R_{\phi-\pi/2}^{\pi/2}R_z^{-\pi/2}R_{\phi+\pi/2}^{\pi/2}\nonumber\\
&=&\id\otimes R_{\phi}^{\pi/2}
\ee

\begin{figure}[htb]
\includegraphics[scale=0.35]{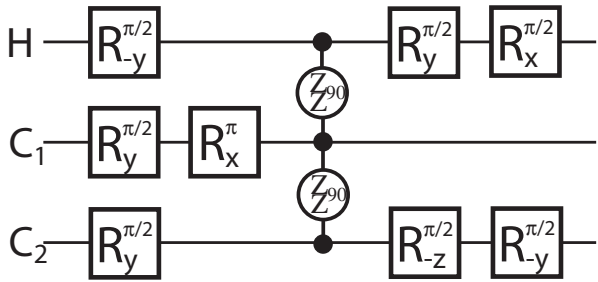}
\caption{Circuit used to implement one step of the DTQW on the TCE.  The z-rotation on $C_2$ occurs  naturally during the coupling with $C_1$.  Note also that the refocusing pulses are not shown in this pulse sequence.     \label{tcecircuit}}
\end{figure}

Similar pulse sequences can be derived to perform a $\pi$ rotation on $C_2$ and selective pulses on $C_1$.  Because of the different form of selective pulses used, the pulse sequences were written and optimized by hand.  This required a different pulse sequence implementation of the quantum walk unitary which avoided as much as possible selective pulses and $z$-rotations where possible.  The one $z$-rotation used, is a natural outcome of the $C_1-C_2$ coupling gate as described above.   This alternative pulse sequence is shown in Fig. \ref{tcecircuit}.

\subsubsection{Experimental Results}

Fidelity results, similar to those calculated for the crotonic acid experiments are shown in Table \ref{resultstce}.   Clearly this experiment was not as successful as the implementation on the crotonic acid molecule. There are two main reasons for this loss of fidelity.  Firstly, the chemical shift difference between the two carbons is very small.  Because of this, the secular approximation no longer holds and thus the coupling between the two carbon spins can no longer be approximated by the Ising form $Z_{C_1}\otimes Z_{C_2}$.  Indeed, it has to take all the strong coupling terms into account, i.e $\vec{S}_{C_1}\cdot\vec{S}_{C_2}=X_{C_1}\otimes X_{C_2}+Y_{C_1}\otimes Y_{C_2}+Z_{C_1}\otimes Z_{C_2} $. 

 \begin{table}
 \begin{tabular}{||c|c|c||}
 \hline\hline
 &Experimental & Simulated  \\
 \hline\hline
 Step 0  & $98\pm6$ & ---\\
 \hline
Step 1 & $85\pm5$ &96\\
\hline
Step 2 & $82\pm4$ &94\\
\hline
Step 3 & $70\pm4$ &93\\
\hline
Step 4 & $80\pm4$ &90\\
\hline
Step 5 & $76\pm4$ &89\\
\hline
Step 6 & $65\pm4$ &86\\
\hline
Step 7 & $53\pm4$ &84\\
\hline
Step 8 & $43\pm4$ &83\\
 \hline\hline
 \end{tabular}
 \caption{Experimental and simulated fidelities (in percent) for the implementation of 8 steps of the DTQW on the molecule TCE.  Again, simulation of the pseudo-pure state was not performed.  }\label{resultstce}
 \end{table}
 
Unfortunately, this strong coupling renders our ideal  $ZZ$ gates much less precise. Every coupling gate performed added XX and YY error terms which we could not refocus.  This coupling also caused problems during our selective carbon rotations.  Although the coupling is small, there is an unrefocusable coupling of $\frac{\pi J_{C_1C_2}}{4\Delta\nu}\approx 4.27^{\circ}$. Our only way to minimize these errors was to optimize the delay times analytically and from numerical simulations.   However, these did not correspond well to the experimentally determined optimal values.  This point also clearly demonstrates the second reason for the less satisfactory results on TCE.  We were unable to use the numerical optimization of the pulse sequence compiler used for crotonic acid.  The compiler provides a systematic and reliable way to produce pulse sequences which implement unitaries with high fidelity and is clearly superior to writing and optimizing pulse sequences by hand.  These experiments also showed the limits of our simulator.  For the crotonic acid experiments, where only soft pulses were used, the r.f. power applied changed slowly and the simulator was faithful to what r.f. power the spins were experiencing.  In TCE, where control was achieved only through short hard pulses, other effects such as phase transients enter and the spins might experience an r.f. field much different from the ideal square pulse simulated.  To fully understand the issues surrounding hard pulse control a much more detailed study of the probe response must be undertaken.  This underlines a key point:  control of a more complex and strongly coupled system could be obtained through sophisticated control techniques such as strongly modulating pulses \cite{FPB+02a} ; however, it seems prudent to invest the effort in a wise choice of molecule.   

\section{Conclusion}
We have presented the first experimental implementation of a coined discrete time quantum walk.  It showed a clear difference with the classical coined quantum walk, since the DTQW  possesses   destructive interference and periodicity in its evolution.  A proof of principle like this lays  down the path to more elaborate experiments using discrete quantum walks, such as the database search, walks on hypercube or N-nodes circle or a more profound study of the effect of decoherence on the walk.  This paper also demonstrates the importance of choosing a natural Hamiltonian well suited to automated control in the context of quantum information processing.  

\begin{acknowledgments}
C. R, M. L, J.-C. B. and R. L. would like to thank M. J. Ditty for technical support with the work on the spectrometer and C. Negrevergne for helpful discussion.  This work has been supported by NSERC, ARDA and CFI. 

\end{acknowledgments}

\end{document}